\documentclass[twocolumn]{aastex63}
\pdfoutput=1 
\usepackage{amsmath,amstext}
\usepackage[T1]{fontenc}
\usepackage[figure,figure*]{hypcap}
\usepackage{graphicx}
\usepackage{subfigure}
\usepackage{tabularx}
\usepackage{needspace,enumitem}

\newcommand{\at}{AT\,2019dsg}

\shorttitle{AT2019dsg}
\shortauthors{Cendes et al.}

\begin{document}

\title{Radio Observations of an Ordinary Outflow from the Tidal Disruption Event AT2019dsg}

\correspondingauthor{Yvette Cendes}
\email{yvette.cendes@cfa.harvard.edu}

\author[0000-0001-7007-6295]{Y. Cendes}
\affiliation{Center for Astrophysics | Harvard \& Smithsonian,
Cambridge, MA 02138, USA}

\author[0000-0002-8297-2473]{K. D. Alexander}\altaffiliation{NHFP Einstein Fellow}
\affiliation{Center for Interdisciplinary Exploration and Research in Astrophysics (CIERA) and Department of Physics and Astronomy, Northwestern University, Evanston, IL 60208, USA}

\author{E. Berger}
\affiliation{Center for Astrophysics | Harvard \& Smithsonian,
Cambridge, MA 02138, USA}

\author{T. Eftekhari}
\affiliation{Center for Astrophysics | Harvard \& Smithsonian,
Cambridge, MA 02138, USA}

\author[0000-0003-3734-3587]{P. K. G. Williams}
\affiliation{Center for Astrophysics | Harvard \& Smithsonian,
Cambridge, MA 02138, USA}
\affiliation{American Astronomical Society, 1667 K Street NW, Suite 800
Washington, DC 20006 USA}

\author[0000-0002-7706-5668]{R.~Chornock}
\affiliation{Center for Interdisciplinary Exploration and Research in Astrophysics (CIERA) and Department of Physics and Astronomy, Northwestern University, Evanston, IL 60208, USA}

\begin{abstract}

We present detailed radio observations of the tidal disruption event (TDE) AT2019dsg, obtained with the Very Large Array (VLA) and the Atacama Large Millimeter/submillimeter Array (ALMA), and spanning $55-560$ days post-disruption.  We find that the peak brightness of the radio emission increases until $\sim 200$ days and subsequently begins to decrease steadily.  Using the standard equipartition analysis, including the effects of synchrotron cooling as determined by the joint VLA-ALMA spectral energy distributions, we find that the outflow powering the radio emission is in roughly free expansion with a velocity of $\approx 0.07c$, while its kinetic energy increases by a factor of about 5 from 55 to 200 days and plateaus at $\approx 4.4\times 10^{48}$ erg thereafter.  The ambient density traced by the outflow declines as $\approx R^{-1.7}$ on a scale of $\approx (1-4)\times 10^{16}$ cm ($\approx 6300-25000$ $R_s$), followed by a steeper decline to $\approx 7\times 10^{16}$ cm ($\approx 44000$ $R_s$).  Allowing for a collimated geometry, we find that to reach even mildly relativistic velocities ($\Gamma=2$) the outflow requires an opening angle of $\theta_j\approx 2^\circ$, which is narrow even by the standards of GRB jets; a truly relativistic outflow requires an unphysically narrow jet. The outflow velocity and kinetic energy in AT2019dsg are typical of previous non-relativistic TDEs, and comparable to those from Type Ib/c supernovae, raising doubts about the claimed association with a high-energy neutrino event. 

\end{abstract}


\section{Introduction} 
\label{sec:intro}

A tidal disruption event (TDE) occurs when a star wanders sufficiently close to a supermassive black hole (SMBH) to be torn apart by tidal forces.  In recent years, the number of observed TDEs has increased dramatically, primarily thanks to wide-field optical time-domain surveys\footnote{See http://tde.space.}.  So far, only about 10 TDEs (about 10\% of the known sample) have been detected in the radio during dedicated follow-up observations \citep[see Table 1; ][]{Alexander2020}.  Even within this small sample it appears that TDEs launch either non-relativistic outflows ($v\sim 0.1c$) with an energy scale of $E_K\sim 10^{48}-10^{49}$ erg (e.g., \citealt{Alexander2016}), or much more rarely relativistic outflows as observed in Swift J1644+57 with $\Gamma\sim {\rm few}$ and $E_K\sim 10^{52}$ erg (e.g., \citealt{Zauderer2011,p1}).   The origin of the outflowing material, its relation to the overall TDE properties, and the physical distinction between events that launch relativistic jets and non-relativistic outflows remain a matter of debate. 

The TDE \at\ ($z = 0.051$) was discovered by the Zwicky Transient Facility (ZTF) on 2019 April 9, and classified as a TDE based on its optical spectrum.  Its peak optical luminosity ($\sim10^{44.5}$ erg s$^{-1}$) places it in the top 10\% of optical TDEs to date \citep{vanVelzen2020}, and a high level of optical polarization was observed at early times \citep{Lee2020}.  \at\ also exhibited X-ray emission and radio emission in the first few months following discovery \citep{Cannizzaro2020,Stein2020}.  Additionally, \citet{Stein2020} claim a potential coincident high-energy neutrino with the spatial location of \at, but several months after discovery (on 2019 October 1); the emission mechanism for such a neutrino is debated \citep{Fang2020,Winter2020,Liu2020,Murase2020}.

Here, we present multi-frequency radio observations of \at\ on a timescale of about 55 to 560 days post disruption, which we use to infer the time evolution of the TDE outflow's energy and velocity, as well as the circumnuclear medium density profile.  In \S\ref{sec:obs}, we present our observations using both the Karl G.~Jansky Very Large Array (VLA) and the Atacama Large Millimeter/submillimeter Array (ALMA).  In \S\ref{sec:modeling} we model the individual radio spectral energy distributions (SEDs) and carry out an equipartition analysis to determine the physical properties of the outflow and the ambient medium.  In \S\ref{sec:discussion} we discuss our findings in the context of the TDE population.  We summarize our conclusions in \S\ref{sec:conclusions}.

\section{Observations}
\label{sec:obs}

We obtained radio observations of \at\ with the VLA spanning from L- to K-band ($1-26.5$ GHz; Program IDs: 19A-013 and 20A-372; PI: Alexander).  The data are summarized in Table~\ref{tab:obs}. We processed the data using standard data reduction procedures in the Common Astronomy Software Application package (CASA; \citealt{McMullin2007}) accessed through the python-based \texttt{pwkit} package\footnote{https://github.com/pkgw/pwkit} \citep{Williams2017}.  We performed bandpass and flux density calibration using either 3C286 or 3C147 as the primary calibrator for all observations and frequencies. We used J2035+1056 as the phase calibrator for L and S bands and J2049+1003 as the phase calibrator for all other frequencies. We imaged the data using the CASA task CLEAN, splitting the data into subbands by frequency when the target was sufficiently bright. We obtained all flux densities and uncertainties using the {\tt imtool fitsrc} command within {\tt pwkit}. We assumed a point source fit, as preferred by the data.

We also observed \at\ with ALMA in band 3 (mean frequency of 97.5 GHz) on 2019 June 22 and September 17 (Table~\ref{tab:obs}).  For the ALMA observations, we used the standard NRAO pipeline to calibrate and image the data. The source was not bright enough for self-calibration. We detect \at\ in the June observation and derive an upper limit in the September observation (Table \ref{tab:obs}). 

\startlongtable
\begin{deluxetable}{lcccc}
\tablecolumns{4}
\tablecaption{VLA and ALMA Observations of \at}
\tablehead{
Date       &
$\delta t$ &  
Array Con- &
$\nu$      & 
$F_\nu$    \\
(UTC) & 
(d)   & 
figuration &
(GHz) & 
(mJy)
}
\startdata
2019 May 24 & 55 & B & 5 & $0.09\pm 0.01$ \\
& &    & 7 & $0.19\pm 0.01$ \\
& &    & 13 & $0.48\pm 0.02$ \\
& &    & 15 & $0.54\pm 0.02$ \\
& &   & 17 & $0.58\pm 0.02$ \\\hline
2019 May 29 & 60 & B & 3.4 & $0.77\pm 0.03$ \\
& &    & 9 & $0.43\pm 0.03$ \\
& &    & 11 & $0.51\pm 0.02$ \\
& &    & 13 & $0.63\pm 0.04$ \\
& &    & 15 & $0.61\pm 0.03$ \\
& &    & 17 & $0.68\pm 0.03$ \\
& &    & 19 & $0.73\pm 0.04$ \\
& &    & 21 & $0.71\pm 0.05$ \\
& &    & 23 & $0.67\pm 0.04$ \\
& &    & 25 & $0.59\pm 0.04$ \\
& &    & 30 & $0.59\pm 0.04$ \\
& &    & 32 & $0.48\pm 0.04$ \\
& &    & 34 & $0.51\pm 0.08$ \\
& &    & 36 & $0.46\pm 0.05$ \\\hline
2019 June 20 & 82 & B & 5 & $0.26\pm 0.02$ \\
& &    & 7 & $0.49\pm 0.02$ \\
& &    & 9 & $0.68\pm 0.02$ \\
& &    & 11 & $0.72\pm 0.03$ \\
& &    & 13 & $0.75\pm 0.02$ \\
& &    & 15 & $0.65\pm 0.03$ \\
& &    & 17 & $0.63\pm 0.02$ \\
& &    & 19 & $0.51\pm 0.03$ \\
& &    & 21 & $0.46\pm 0.04$ \\
& &    & 23 & $0.43\pm 0.05$ \\
& &    & 25 & $0.40\pm 0.03$ \\
& &    & 30 & $0.33\pm 0.04$ \\
& &    & 32 & $0.27\pm 0.03$ \\
& &    & 34 & $0.31\pm 0.03$ \\
& &    & 36 & $0.24\pm 0.04$ \\\hline
2019 June 22 & 84 & C43-9/10 & 97.5& $0.07\pm 0.01$\\\hline
2019 Sept 7 & 161 & A & 2.6 & $0.29\pm 0.03$ \\
& &    & 3.4 & $0.42\pm 0.03$ \\
& &    & 5 & $0.56\pm 0.02$ \\
& &    & 7 & $0.89\pm 0.02$ \\
& &    & 9 & $1.03\pm 0.03$ \\
& &    & 11 & $1.08\pm 0.05$ \\
& &    & 13 & $1.02\pm 0.02$ \\
& &    & 15 & $0.91\pm 0.02$ \\
& &    & 17 & $0.80\pm 0.04$ \\
& &    & 19 & $0.70\pm 0.04$ \\
& &    & 21 & $0.62\pm 0.05$ \\
& &    & 23 & $0.58\pm 0.06$ \\
& &    & 25 & $0.41\pm 0.05$ \\\hline
2019 Sept 17 & 188 & C43-6 & 97.5 & $<0.09$\\\hline
2020 Jan 24 & 300 & D & 2.6 & $0.67\pm 0.04$ \\
& &    & 3.4 & $0.82\pm 0.03$ \\
& &    & 5 & $0.79\pm 0.03$ \\
& &    & 7 & $0.68\pm 0.03$ \\
& &    & 9 & $0.46\pm 0.03$ \\
& &    & 11 & $0.48\pm 0.45$ \\
& &    & 13 & $0.31\pm 0.02$ \\
& &    & 15 & $0.27\pm 0.02$ \\
& &    & 17 & $0.23\pm 0.03$ \\\hline
2020 Oct 11 & 561 & B & 1.5 & $0.35\pm 0.05$ \\
& &    & 2.6 & $0.26\pm 0.02$ \\
& &    & 3.4 & $0.17\pm 0.01$ \\
& &    & 6.0 & $0.13\pm 0.01$ \\\hline
\enddata
\tablecomments{Errors are statistical only.  Upper limits are $3\sigma$.}
\label{tab:obs}
\end{deluxetable}

\begin{figure*}
    \includegraphics[width=.7\columnwidth]{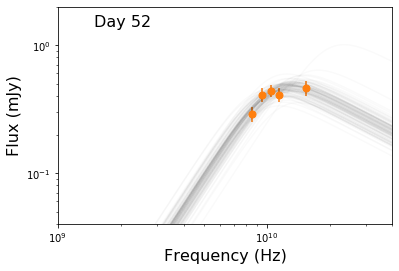}
    \includegraphics[width=.7\columnwidth]{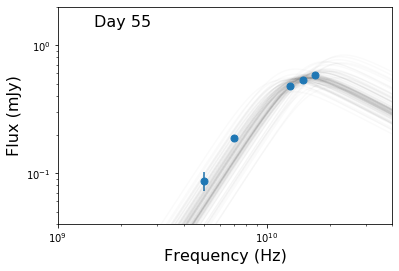}
    \includegraphics[width=.7\columnwidth]{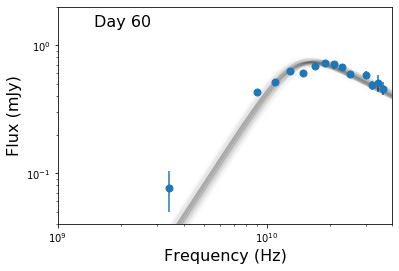}
    \includegraphics[width=.7\columnwidth]{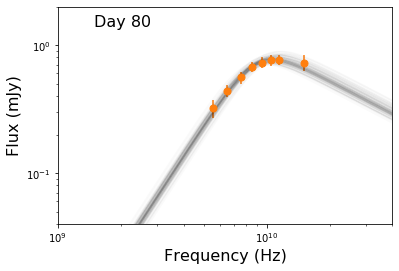}
    \includegraphics[width=.7\columnwidth]{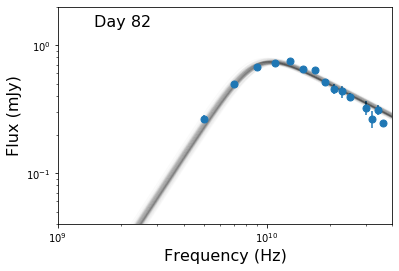}
    \includegraphics[width=.7\columnwidth]{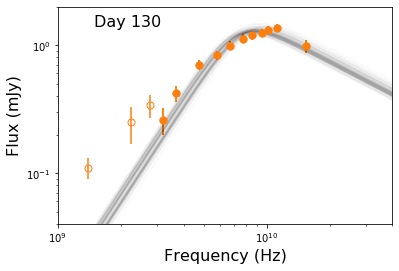}
    \includegraphics[width=.7\columnwidth]{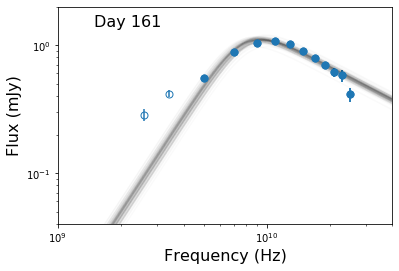}
    \includegraphics[width=.7\columnwidth]{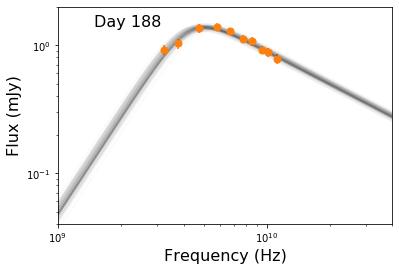}
    \includegraphics[width=.7\columnwidth]{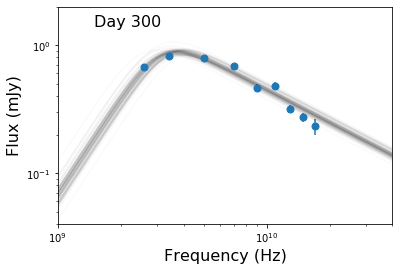}
    \includegraphics[width=.7\columnwidth]{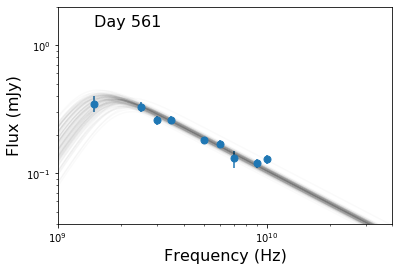}
    \label{fig:sed}
    \caption{The radio spectral energy distributions for our VLA data (blue), as well as the data in \citet{Stein2020} (orange).  The open circles indicate detections that are excluded from the SED fitting. Statistical uncertainties are included in all points but are too small to be visible in most cases.  The grey lines are representative fits from our MCMC modeling (\S\ref{sec:sed}).}
\end{figure*}

\section{Modeling and Analysis}
\label{sec:modeling}

To roughly determine the time that the radio-emitting outflow was launched, we fit a second-order polynomial to the first 5 optical $r$-band flux measurements, which capture the rising part of the light curve \citep[see Table S6; ][]{Stein2020} in order to estimate the time of zero optical flux as a proxy for the time of disruption.  We find a date of 2019 March 30 (MJD 58572.8; about 10 days before optical discovery), which we use in Table~\ref{tab:obs} and in our subsequent analysis.  With this choice we find that the outflow is roughly in free expansion (\S\ref{sec:equi-results}), validating our approach.

\subsection{Modeling of the Radio Spectral Energy Distributions}
\label{sec:sed}

The individual radio SEDs are shown in Figure~\ref{fig:sed}, where we have also included for completeness\footnote{We compared our VLA data to the e-MERLIN observations at 5.1 GHz, some of which overlap with our observations \citep{Cannizzaro2020}.  The September 7 e-MERLIN flux densities appear systematically higher than our own.  After further investigation, we conclude that this is due to differences in the data processing technique.}, and therefore do not incorporate the e-MERLIN data in our SEDs. the data presented in \citet{Stein2020}.  The SEDs are characteristic of self-absorbed synchrotron emission, with a well defined peak frequency ($\nu_p$) and flux density ($F_{\nu,p}$), and a spectral shape of $F_\nu\approx \nu^{5/2}$ below $\nu_p$.  

We fit the SEDs\footnote{For the SEDs at days 130 and 161 we exclude from the modeling the data at $1-3$ GHz that clearly deviate from the expected spectral shape (open points in Figure~\ref{fig:sed}).  We attribute these fluctuations to interstellar scintillation (\S\ \ref{sec:scintillation}).} with the model developed by \citet{Granot2002} for synchrotron emission from gamma-ray burst (GRB) afterglows, specifically in the regime where $\nu_{m}\ll\nu_{a}$; this is relevant for non-relativistic sources as validated by the analysis below.  We have chosen the stellar wind (k=2) solutions from \citet{Granot2002} as the closest approximation to the expected circumnuclear density profile surrounding the TDE SMBH.  The model SED is given by:
\begin{multline}
\indent F_\nu = F_\nu (\nu_m) \Bigg[ \Big(\frac{\nu}{\nu_m}\Big)^2 e^{-s_4 (\nu/\nu_m)^{2/3}}+\Big(\frac{\nu}{\nu_m}\Big)^{5/2}\Bigg] \ 
\\
\times \Bigg[1+\Big(\frac{\nu}{\nu_a}\Big)^{s_2 (\beta_2-\beta_3)}\Bigg]^{-1/s_2},
\label{eq:second-spec}
\end{multline}
where $\beta_2 = 5/2$, $\beta_3 = (1-p)/2$, $s_4 = 3.63p - 1.60$, and $s_2 = 1.25 - 0.18p$.  Here, $p$ is the electron energy distribution power law index --- $N(\gamma_e)\propto \gamma_e^{-p}$ for $\gamma_e\ge\gamma_m$ --- $\nu_m$ is the frequency corresponding to $\gamma_m$, $\nu_a$ is the synchrotron self-absorption frequency, and $F_\nu(\nu_m)$ is the flux normalization at $\nu=\nu_m$. 

We determine the best fit parameters using the Python Markov Chain Monte Carlo (MCMC) module \texttt{emcee} \citep{Foreman-Mackey2013}, assuming a Gaussian likelihood for the parameters $F_{\nu}(\nu_m)$ and $\nu_a$.  Since $\nu_m\ll\nu_a$ we set its value below the range of our data ($\nu_m=0.1$ GHz).  In an initial round of modeling, we first fit for $p$ as a free parameter in each epoch (with a uniform prior of $p=2-4$).  We then exclude the epochs where the resulting uncertainty on $p$ is $\delta p\ge 0.2p$, which is due to the paucity of data at $\gtrsim \nu_a$ (e.g., SEDs at days 52, 55).  We find no evidence for a time evolution in the value of $p$, with a weighted average of $p=2.7\pm 0.2$, which we adopt in our subsequent analysis.  We also include a parameter that accounts for additional systematic uncertainty beyond the statistical uncertainty on the individual data points.  The posterior distributions are sampled using 100 MCMC chains, which were run for 2,000 steps, discarding the first 1,000 steps to ensure the samples have sufficiently converged by examining the sampler distribution.

\begin{figure}
    \includegraphics[width=.95\columnwidth]{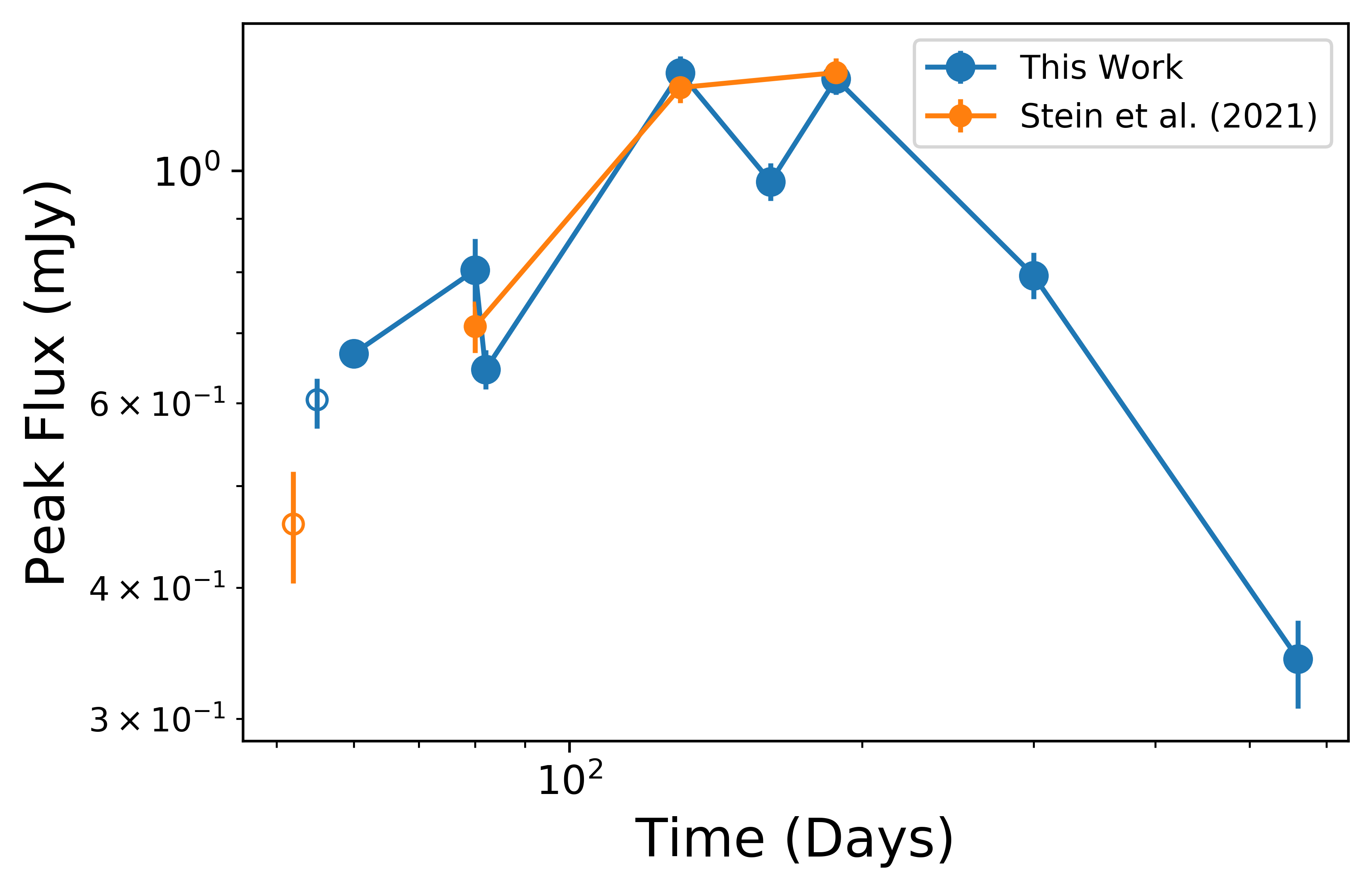}
    \label{fig:F_p}
    \includegraphics[width=.95\columnwidth]{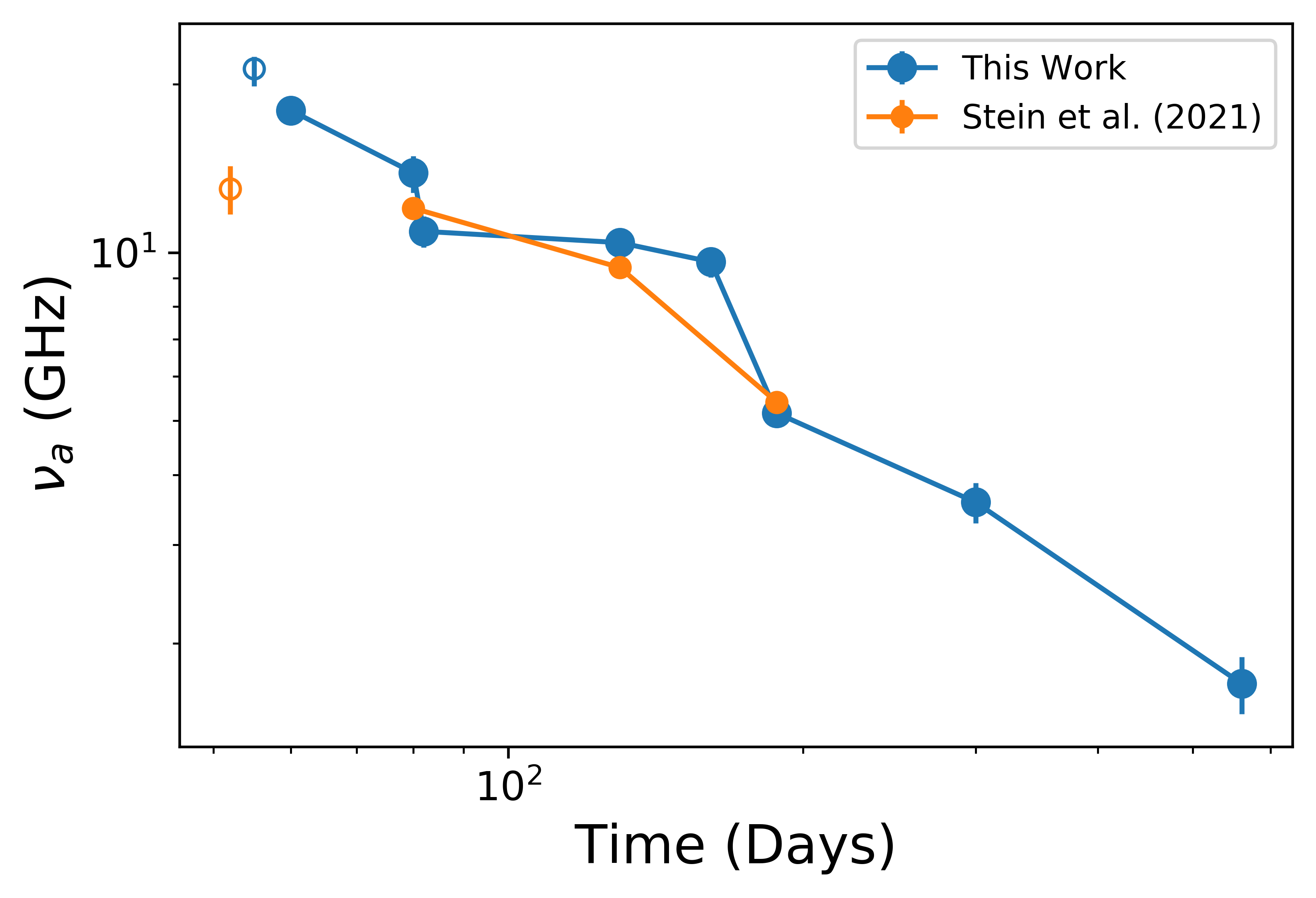}
    \label{fig:nu-a}
    \caption{Time evolution of the synchrotron peak flux density ({\it Left}) and peak frequency ({\it Right}) from the SED modeling of the data in Figure~\ref{fig:sed}.  The blue points indicate our fits to the entire data set (including the data from \citealt{Stein2020}), while the orange points are taken from the analysis of \citet{Stein2020}.  Open circles represent epochs in which the values should be taken as lower limits due to the lack of data above the peak frequency. These points are not used in the subsequent analysis.}
\end{figure}

From the SED fits we determine the frequency and flux density of the model SED peak, $\nu_p$ and $F_p$, respectively.  The time evolution of these values is shown in Figure~\ref{fig:F_p}, where we also include for comparison the values reported by \citet{Stein2020}, although these authors assume $p=3$ in their analysis.  The parameter values are listed in Table~\ref{tab:SED-fits}.  In the first two available epochs, 52 days from \citet{Stein2020} and 55 days from our data, we consider $\nu_p$ and $F_p$ as essentially lower limits since the SED peak is not well captured by the data.  We find that $\nu_p$ declines steadily at $\approx 60-560$ days as about $t^{-0.85}$.  The evolution of $F_p$ is more complex and less typical, with an initial rise by about a factor of $3-4$ to about 200 days, and a subsequent decline by about a factor of 4 to 560 days.

\begin{deluxetable*}{lcc|cccccr}
\label{tab:SED-fits}
\tablecolumns{8}
\tablecaption{Spectral Energy Distribution and Equipartition Model Parameters}
\tablehead{
	\colhead{$\delta t$} &
	\colhead{$F_p$} &
	\colhead{log($\nu_p$)} &
	\colhead{log($R_{\rm eq}$)} &
	\colhead{log($E_{\rm eq}$)} &
	\colhead{log($B$)} &
	\colhead{log($N_e$)} &
	\colhead{log($n_{\rm ext}$)} &
	\colhead{log($v_{\rm eq}$)}
	\\
    \colhead{(d)} &
    \colhead{(mJy)} &
    \colhead{(Hz)} &
    \colhead{(cm)} &
    \colhead{(erg)} &
    \colhead{(G)} &
    &
    \colhead{(cm$^{-3}$)} &
    \colhead{(m/s)}
}
 \startdata
 52$^{\dagger*}$&$0.47\substack{+0.06 \\ -0.09}$&$10.20\substack{+0.10 \\ -0.09}$&$15.92\substack{+0.1 \\ -0.05}$&$47.66\substack{+0.10 \\ -0.10}$&$0.23\substack{+0.14 \\ -0.17}$&$52.35\substack{+0.10 \\ -0.10}$&$3.93\substack{+0.26 \\ -0.38}$&$7.27\substack{+0.10 \\ -0.13}$\\
 55$^\dagger$&$0.60\substack{+0.03 \\ -0.04}$&$10.32\substack{+0.13 \\ -0.09}$&$15.84\substack{+0.09 \\ -0.09}$&$47.66\substack{+0.03 \\ -0.03}$&$0.34\substack{+0.05 \\ -0.06}$&$52.35\substack{+0.03 \\ -0.03}$&$4.15\substack{+0.10 \\ -0.12}$&$7.17\substack{+0.04 \\ -0.04}$\\
 60&$0.67\substack{+0.01 \\ -0.01}$&$10.26\substack{+0.03 \\ -0.03}$&$15.94\substack{+0.03 \\ -0.03}$&$47.78\substack{+0.06 \\ -0.03}$&$0.26\substack{+0.03 \\ -0.03}$&$52.47\substack{+0.03 \\ -0.04}$&$3.99\substack{+0.06 \\ -0.06}$&$7.23\substack{+0.03 \\ -0.03}$\\
 80$^*$&$0.80\substack{+0.06 \\ -0.06}$&$10.14\substack{+0.05 \\ -0.06}$&$16.09\substack{+0.04 \\ -0.04}$&$47.99\substack{+0.05 \\ -0.05}$&$0.14\substack{+0.07 \\ -0.05}$&$52.68\substack{+0.05 \\ -0.05}$&$3.75\substack{+0.13 \\ -0.15}$&$7.25\substack{+0.05 \\ -0.06}$\\
 82&$0.65\substack{+0.03 \\ -0.03}$&$10.03\substack{+0.03 \\ -0.03}$&$16.15\substack{+0.07 \\ -0.07}$&$47.98\substack{+0.09 \\ -0.10}$&$0.05\substack{+0.03 \\ -0.07}$&$52.67\substack{+0.09 \\ -0.10}$&$3.56\substack{+0.12 \\ -0.14}$&$7.30\substack{+0.07 \\ -0.07}$\\
 130$^*$&$1.24\substack{+0.05 \\ -0.05}$&$10.01\substack{+0.04 \\ -0.04}$&$16.30\substack{+0.04 \\ -0.04}$&$48.34\substack{+0.04 \\ -0.04}$&$-0.00\substack{+0.04 \\ -0.05}$&$53.03\substack{+0.04 \\ -0.04}$&$3.46\substack{+0.08 \\ -0.09}$&$7.25\substack{+0.04 \\ -0.04}$\\
 161&$0.98\substack{+0.04 \\ -0.04}$&$9.98\substack{+0.04 \\ -0.03}$&$16.29\substack{+0.07 \\ -0.07}$&$48.25\substack{+0.09 \\ -0.10}$&$-0.03\substack{+0.06 \\ -0.07}$&$52.94\substack{+0.09 \\ -0.10}$&$3.41\substack{+0.12 \\ -0.14}$&$7.15\substack{+0.07 \\ -0.07}$\\	
188$^*$&$1.22\substack{+0.04 \\ -0.04}$&$9.71\substack{+0.04\\ -0.02}$&$16.60\substack{+0.06 \\ -0.06}$&$48.64\substack{+0.05 \\ -0.08}$&$-0.30\substack{+0.02 \\ -0.04}$&$53.33\substack{+0.07 \\ -0.08}$&$2.86\substack{+0.10 \\ -0.11}$&$7.39\substack{+0.06 \\ -0.06}$\\
300&$0.79\substack{+0.04 \\ -0.04}$&$9.55\substack{+0.03 \\ -0.05}$&$16.68\substack{+0.1 \\ -0.11}$&$48.58\substack{+0.12 \\ -0.14}$&$-0.45\substack{+0.09 \\ -0.09}$&$53.26\substack{+0.12 \\ -0.14}$&$2.57\substack{+0.17 \\ -0.19}$&$7.27\substack{+0.10 \\ -0.11}$\\
561&$0.34\substack{+0.03 \\ -0.04}$&$9.23\substack{+0.11 \\ -0.11}$&$16.85\substack{+0.18 \\ -0.24}$&$48.48\substack{+0.23 \\ -0.38}$&$-0.76\substack{+0.14 \\ -0.19}$&$53.17\substack{+0.23 \\ -0.038}$&$1.95\substack{+0.26 \\ -0.42}$&$7.16\substack{+0.18 \\ -0.25}$\\	
 \enddata
 \tablecomments{Values are assuming $f_A = 1$, $f_V = 0.36$, $\epsilon_e = 0.1$, and $\epsilon_b = 0.02$.\\ $^*$ Flux density measurements obtained from \citet{Stein2020}.\\ $^\dagger$ Parameters derived from these observations are excluded from our analysis, but included here for completeness (see text).}
 \end{deluxetable*}

\subsection{Equipartition Analysis}
\label{sec:equi}

Using the inferred values of $\nu_p$ and $F_p$ from \S\ref{sec:sed}, we can now derive the physical properties of the outflow using an equipartition analysis.  We focus on the case of a non-relativistic spherical outflow using the following expressions for the radius and kinetic energy \citep[see Equations 27 and 28 in ][]{Duran2013}, with $p=2.7$:
\begin{multline} 
R_{\rm eq}\approx (9.1\times10^{22}\,\, \textrm{cm})\times F_{p,\rm mJy}^{8.7/18.4} d_{L,28}^{17.4/18.4} \nu_{p,10}^{-1}
\\
\times(1+z)^{-27.1/18.4} f_A^{-7.7/18.4} f_V^{-1/18.4} 4^{1/18.4} \epsilon^{1/17}
\\
\times \xi^{1/18.4} \gamma_{m}^{-0.7/18.4}
\label{eq:rad}
\end{multline}
\begin{multline} 
E_{\rm eq} \approx (2.17\times10^{50}\,\, \textrm{erg})\times F_{p,\rm mJy}^{22.1/18.4} d_{L,28}^{44.2/18.4} \nu_{p,10}^{-1}
\\
\times (1+z)^{-40.5/18.4} f_A^{-11.1/18.4} f_V^{7.4/18.4} 4^{11/18.4} \xi^{11/18.4}
\\ \times\gamma_m^{0.75} [(11/17)\epsilon^{-6/17}+(6/17)\epsilon^{11/17}],
\label{eq:energy}
 \end{multline}
where $d_L\approx 230$ Mpc is the luminosity distance; $z=0.051$ is the redshift; $f_A=1$ and $f_V= \frac{4}{3}\times (1-0.9^3)=0.36$ are the area and volume filling factors, respectively, where we assume that the emitting region is a shell of thickness $0.1R_{\rm eq}$; and $\gamma_{m}=2$ is the minimum Lorentz factor as relevant for non-relativistic sources \citep{Duran2013}. We chose these factors for $f_A$ and $f_V$ since there is no evidence for significant beaming (see \S\ref{sec:jet}), and thus the simplest assumption is a roughly spherical outflow. The factors of $4^{1/18.4}$ and $4^{11/18.4}$ for the radius and energy, respectively, arise from corrections to the isotropic number of radiating electrons ($N_{e,\rm iso}$) in the non-relativistic case. We further assume that the fraction of post-shock energy in relativistic electrons is $\epsilon_e=0.1$, which leads to correction factors of $\xi^{1/18.4}$ and $\xi^{11/18.4}$ in $R_{\rm eq}$ and $E_{\rm eq}$, respectively, with $\xi = 1 + \epsilon_e^{-1}\approx 11$.  Finally, we parameterize any deviation from equipartition with a correction factor $\epsilon = (11/6)(\epsilon_B/\epsilon_e)$, where $\epsilon_B$ is the fraction of post-shock energy in magnetic fields.

The magnetic field strength and the number of radiating electrons are given by \citep[see Equations 16 and 15 in ][]{Duran2013}:
\begin{multline} 
B \approx (1.3\times10^{-2}\, \textrm{G})\times  F_{p,\rm mJy}^{-2} d_{L,28}^{-4} (1+z)^{7} f_A^{2} R_{\rm eq,17}^{4} \nu_{p,10}^5
\label{eq:mag}
\end{multline}
\begin{multline} 
N_{e} \approx (4\times10^{54})\times F_{p,\rm mJy}^{3} d_{L,28}^{6} (1+z)^{-8} f_A^{-2} R_{\rm eq,17}^{-4} \nu_{p,10}^{-5} \\ \times (\gamma_{a}/\gamma_{m})^{(p-1)},
\label{eq:dens}
\end{multline}
where the Lorentz factor of electrons that radiate at $\nu_a$ is given by:
\begin{multline} 
\gamma_a \approx 525\times F_{p,\rm mJy} d_{L,28}^2 \nu_{p,10}^{-2} (1+z)^{-3} f_A^{-1} R_{\rm eq,17}^{-2},
\label{eq:gamma}
\end{multline}

We will note explicitly that the extra factor of 4 and extra correction factor $(\gamma_{a}/\gamma_{m})^{(p-1)}$ are added correction factors for the Newtonian regime (Duran, private communication). We calculate the density of the ambient medium as $n_{\rm ext} = N_e/4V$, where $V$ is the volume of the emission region assumed to be a spherical shell of thickness $0.1R_{\rm eq}$ and the factor of $\frac{1}{4}$ is due to the shock jump conditions.

\subsection{Cooling Frequency and $\epsilon_{B}$}
\label{sec:cooling}

\begin{figure}[t!]
        \includegraphics[width=.95\columnwidth]{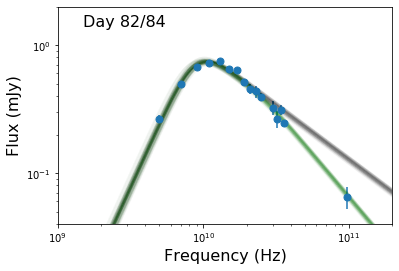}
        \includegraphics[width=.95\columnwidth]{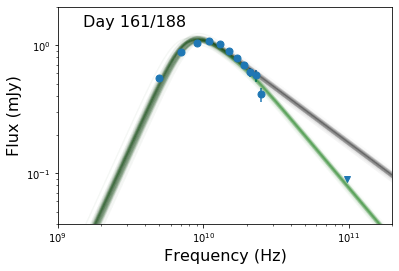}
    \caption{VLA + ALMA data (blue) at $82-84$ days ({\it Top}) and $161-188$ days ({\it Bottom}), where circles are detections and the ALMA upper limit is marked with a triangle.  Also shown are the models from Figure~\ref{fig:sed} (grey; no cooling break), and models that include a synchrotron cooling break (green; Equation~\ref{eq:cooling-component}).  At $82-84$ days we find that $\nu_c\approx 25$ GHz, leading to $\epsilon_B\approx 0.02$ (\S\ref{sec:cooling}).  At $161-188$ days, we find that $\nu_c\approx 20$ GHz, leading to a consistent value of $\epsilon_B\approx  0.01$.}
    \label{fig:cooling}
\end{figure}

Our ALMA observations allow us to investigate the presence of a cooling break between the VLA and ALMA bands.  The synchrotron cooling frequency is given by \citep{Sari1998}:
\begin{equation} 
\nu_c = 2.8\times10^{6} \gamma_c^2 B,
\label{eq:cooling}
\end{equation}
where $\gamma_c = 6 \pi m_e c / \sigma_{T} B^2 t$, with $t$ being the age of the system.

In Figure~\ref{fig:cooling} we show our VLA+ALMA SEDs at $82-84$ and $161-188$ days, along with our model SED from \S\ref{sec:sed}, which does not include a cooling break (grey lines). This model clearly over-predicts the ALMA measurements (and already begins to deviate from the data at $\approx 25-30$ GHz).  The steepening required by the ALMA data is indicative of a cooling break, which we model with an additional multiplicative term to Equation~\ref{eq:second-spec} \citep{Granot2002}:
\begin{equation}
\Bigg[1+\Big(\frac{\nu}{\nu_c}\Big)^{s_3 (\beta_3-\beta_4)}\Bigg]^{-1/s_3},
\label{eq:cooling-component}
\end{equation}
where $\beta_4 = -p/2$ and we use $s_3=10$.\footnote{We note that our choice for $s_3$ is steeper than the value in \citet{Granot2002}, and is motivated by the actual observed sharpness in the break and by the fact that \citet{Granot2002} formalism has been derived in the context of relativistic events (specifically, GRB afterglows).  With potential future detections of cooling breaks in TDEs we may be able to assess whether the shape of the break is similar to this event, or more in line with the \citet{Granot2002} theoretical formalism.}

For our observation at $82-84$ days we find that $\nu_c\approx 25$ GHz provides a much better fit to the high frequency data (Figure~\ref{fig:cooling}).  With the value of $\nu_c$ determined, we adjust the value of $\epsilon_B$ and solve Equation~\ref{eq:cooling} after repeating the equipartition analysis (Equations~\ref{eq:rad} to \ref{eq:dens}) to account for the deviation from equipartition in those parameters.  Specifically, we find that with the inclusion of a cooling frequency, $\epsilon_{B} << 0.1$ in order for these equations to remain consistent, otherwise $\nu_{c}$ is significantly higher than what our data indicates.  With this approach, we find $\epsilon_{B}\approx 0.02$.  For our observations at $161-188$ days, we find that $\nu_c\approx 20$ GHz best fits our data, indicating a consistent value of $\epsilon_B\approx 0.01$.  Since the observation covering $82-84$ days includes a detection, we adopt $\epsilon_{B}\approx 0.02$ in our analysis.  Moreover, using the time evolution of the relevant parameters we find that the time evolution of $\nu_c$ does not violate the non-detection of a break in the other SEDs, which are in any case mainly restricted to $\lesssim 30$ GHz. 

\subsection{Outflow and Ambient Medium Properties}
\label{sec:equi-results}

\begin{figure*}[!ht]
    \includegraphics[width=.95\textwidth]{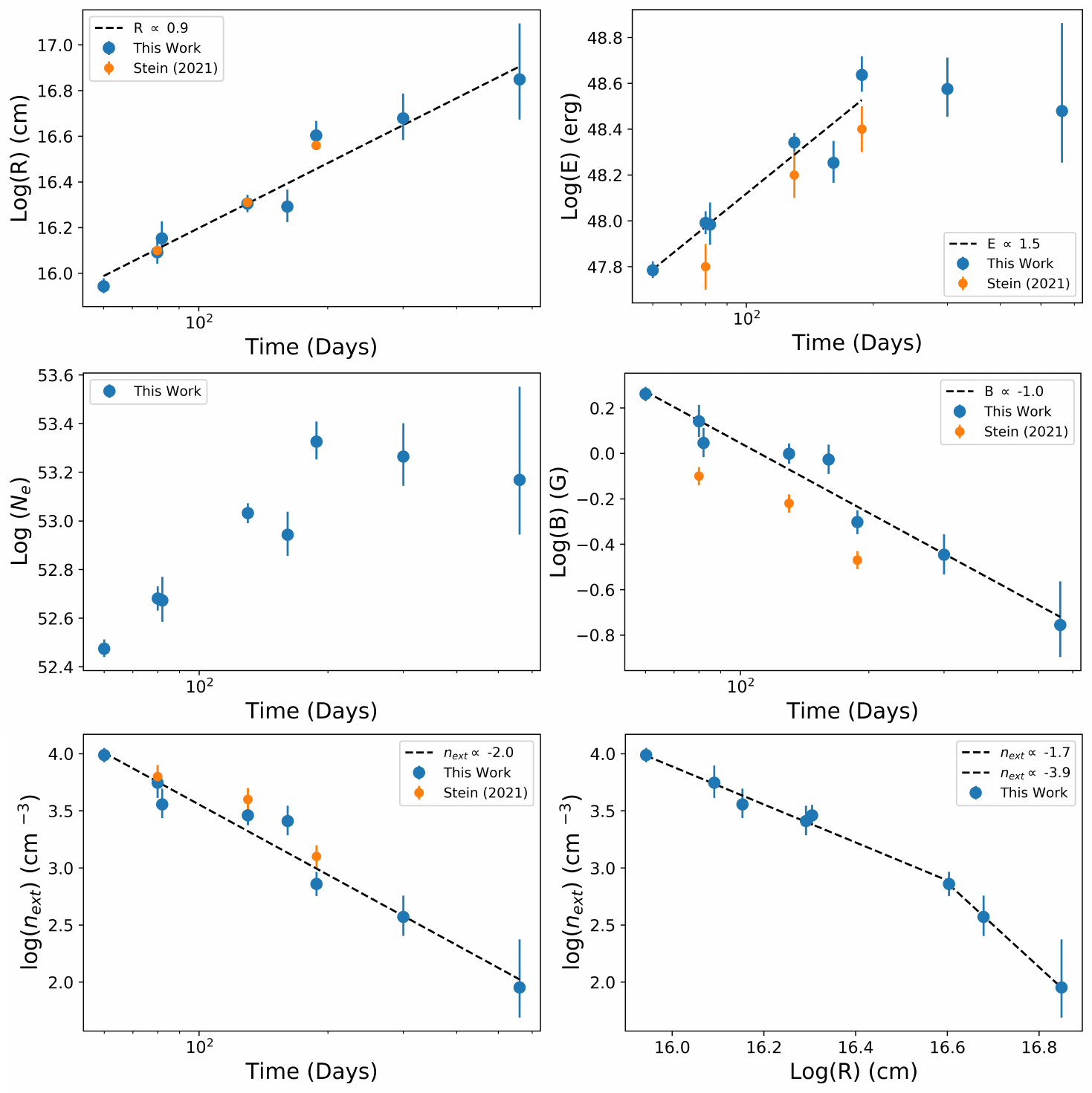}
    \caption{Time evolution of the outflow and density parameters: $R$, $E_K$, $N_e$, $B$, and $n_{\rm ext}$, as well as the radial profile of $n_{\rm ext}$ (bottom right panel).  The blue points mark the results of our analysis of the entire data set (including the data from \citealt{Stein2020}), while the orange points are taken from the analysis of \citet{Stein2020} when assuming a spherical outflow with no deviation from equipartition present.  The dashed lines are power law fits to the data, with the inferred index given in each panel.}
    \label{fig:all-the-things}
\end{figure*}

The inferred outflow parameters as a function of time, as well as the circumnuclear density as a function of radius, are plotted in Figure~\ref{fig:all-the-things}.  The calculated parameters are also listed in Table~\ref{tab:SED-fits}. We find that the radius increases steadily as a power law with $R\propto t^{0.9}$, with a value of about $10^{16}$ cm at 60 days.  This is roughly consistent with free expansion, and we infer a mean velocity of $v\approx 0.07c$ at $60-560$ days, justifying our assumption in \S\ref{sec:equi} of non-relativistic expansion. The kinetic energy exhibits an increase by about a factor of 5 at $60-200$ days ($E_K\propto t^{1.5}$), and then plateaus at a value of $E_K\approx 4\times 10^{48}$ erg.  We additionally find a steady decline in the magnetic field strength, with $B\propto t^{-1.0}$ (or $B\propto R^{-1.1}$), and a value of about $1.8$ G at 60 days. The circumnuclear density evolves as $n_{\rm ext}\propto t^{-2.0}$, or equivalently on average as $\propto R^{-2.1}$.  Plotting $n_{\rm ext}$ as a function of $R$, we find a possibly more complex structure, with a profile of $n_{\rm ext}\propto R^{-1.7}$ at $\approx (1-4)\times 10^{16}$ cm, and a steeper $n_{\rm ext}\propto R^{-3.9}$ to $\approx 7\times 10^{16}$ cm.

The trends in our data are also evident in the more restricted time range of the analysis in \citet{Stein2020} when compared to their equipartition values.  We find systematic offsets in $E_K$ (their values lower by a factor of 1.5), $B$ (lower by a factor of 1.3), and $n_{\rm ext}$ (higher by a factor of 1.4); see Figure~\ref{fig:all-the-things}.  The offsets are due to a combination of a different choice of $p$ (2.7 in our analysis versus 3 in theirs) and our determination of $\epsilon_B\approx 0.02$ based on the detection of a cooling break.

\section{\at\ in the Context of Other TDEs}
\label{sec:discussion}

\subsection{Outflow Velocity and Kinetic Energy}

We find that the radio emission from \at\ is due to a non-relativistic outflow with $v\approx 0.07c$ and a kinetic energy that rises as $E_K\propto t^{1.5}$ at $60-200$ days to a plateau at $4\times 10^{48}$ erg.  The rise in energy could be due to a sustained injection of energy from the central engine (accreting SMBH), or to a rapid initial ejection but with a spread of velocities.  The latter effect is apparent in Type Ib/c SNe and in some long GRBs \citep{Laskar+14,Laskar+15,Margutti2014,Bietenholz2021}. 
In the case of \at\, since the velocity is roughly constant the more likely explanation for the increase in energy is continued injection due to sustained accretion onto the SMBH. A more detailed exploration of this effect will benefit from a detailed analysis of the optical/UV data and an inference of the mass accretion rate as a function of time; this is beyond the scope of this paper.

\begin{figure}[t!]
    \includegraphics[width=1\columnwidth]{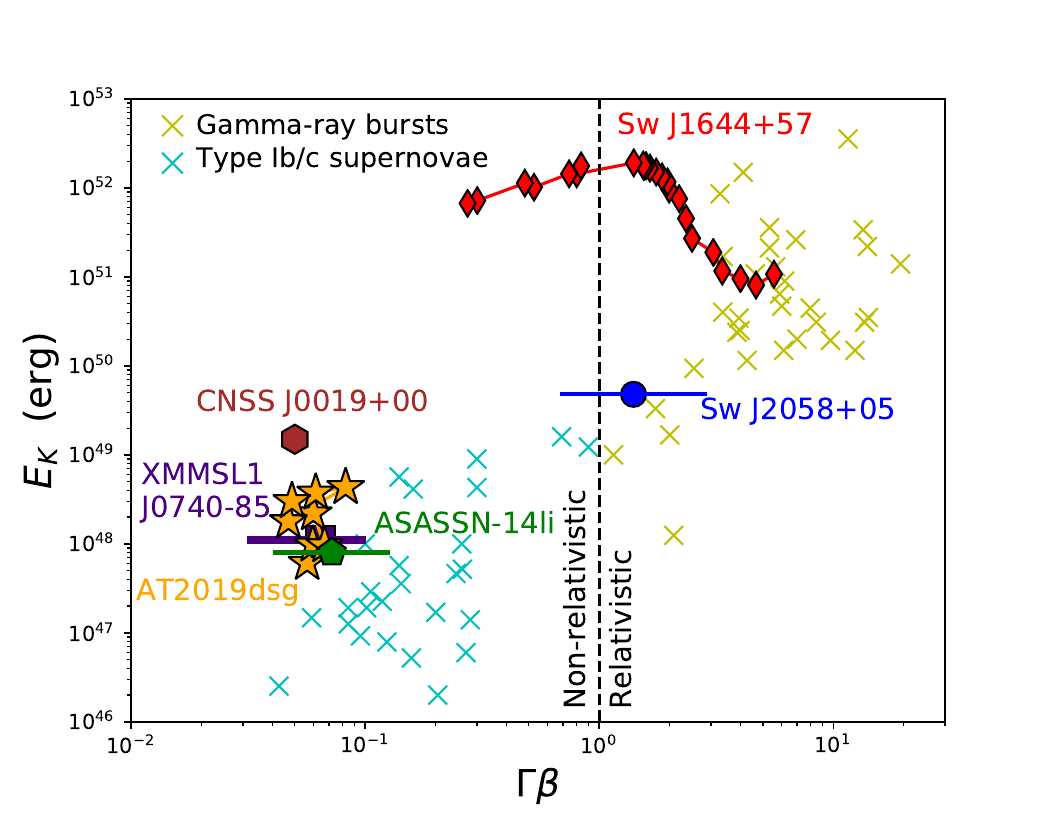}
    \label{fig:energy-tde}
    \caption{The energy/velocity of various TDEs including \at. \at's radio emission shows similar energetics to other transients known to launch non-relativistic outflows, including TDEs \citep{Alexander2016, Alexander2017, Anderson2019} and Type Ib/c supernovae (SNe; sample from \citealt{Margutti2014}). In both the TDE and SN populations, only a small fraction of events launch energetic relativistic jets \citep{Cenko2012,Margutti2014,Zauderer2011,Cendes2021}, suggesting that the conditions required for the formation of such jets are rare.}
\end{figure}

In Figure~\ref{fig:energy-tde} we place the outflow from \at\ in the context of other radio-emitting TDEs, as well as long GRBs and Type Ib/c SNe.  We find that the outflow in \at\ clusters with those of previous TDEs with $E_K\approx 10^{48}-10^{49}$ erg and $v\sim 0.1c$.  It is clearly distinguished from the relativistic TDE Sw J1644+57 with $\Gamma\sim {\rm few}$ and $E_K\sim 10^{52}$ erg \citep{Zauderer2011,p1}. However, we note that like AT2019dsg, Sw J1644+57 also exhibited an order of magnitude increase in its energy at early times \citep{p1}, and to date these TDEs are the only two with multi-frequency radio detections covering the early period post-disruption ($t_d \lesssim 100$ days).

We also find that the outflow energy and velocity of \at\ (and the other non-relativstic TDEs) are comparable to those seen in radio-emitting Type Ib/c SNe (Figure~\ref{fig:energy-tde}).
\citet{Stein2020} report a neutrino event discovered on 2019 October 1, which they associate with \at\ based on a rough spatial coincidence, and despite the several months delay between the disruption and the neutrino event.  A variety of emission mechanisms have been proposed for this neutrino, but many of those that depend on the presence of a jet or outflow in AT2019dsg require a larger outflow energy than we find in our analysis \citep{Fang2020,Winter2020,Liu2020,Murase2020}. The ordinary outflow properties of \at\ in the context of TDEs and Type Ib/c SNe weakens the argument for such an association, as neutrinos have not previously been associated with these events, and Type Ib/c SNe are much more common than TDEs.

\subsection{Relativistic Outflow}
\label{sec:jet}

In our analysis we assume that the outflow from \at\ is isotropic, which is consistent with the inference of a non-relativistic outflow.  If we instead assume that the outflow is collimated, then the inferred radius (and hence the velocity) will increase.  Here we consider what outflow collimation is required to result in a mildly relativistic outflow, with $\Gamma=2$.  This corresponds to the ``narrow jets'' solution in \citet{Duran2013}, with $f_{A} = f_{V} = (\theta_{j}\Gamma)^2$, for a collimated jet with a half-opening angle $\theta_{j}$.  Applying these geometric factors in Equation~\ref{eq:rad}, we find a self-consistent solution for $\theta_{j}\approx 0.035$ rad ($2^{\circ}$), narrow even in the context of GRBs \citep{Frail2001}. To produce a truly relativistic outflow ($\Gamma\sim 10$) would require an unphysical outflow with $\theta_j\ll 1^\circ$).  Therefore, our conclusion that the outflow in \at\ is non-relativistic is robust.

\subsection{Circumnuclear Density}

\begin{figure}
    \includegraphics[width=1\columnwidth]{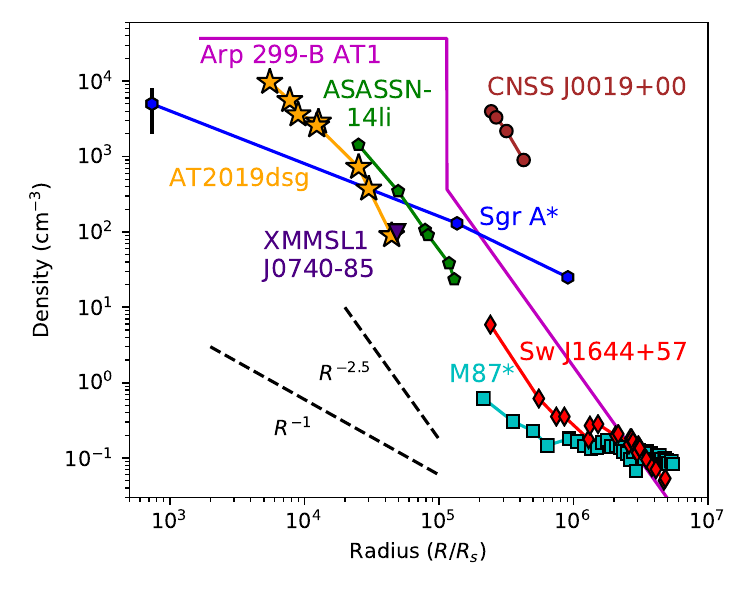}
    \label{fig:dens-tde}
    \caption{The circumnuclear density profile derived from various TDEs including \at, normalized to the Schwarzschild radius of the SMBH at each host galaxy's center. \at's host galaxy has a steep density profile, similar to those seen in other non-relativistic TDEs (e.g.~ASASSN-14li,  \citealt{Alexander2016}, and CNSS J0019+00, \citealt{Anderson2019}). We also show the density profiles calculated for several other TDEs \citep{Alexander2017,Mattila2018,p3} and for the Milky Way \citep{Baganoff2003,Gillessen2019} and M87 \citep{Russell2015}.}
\end{figure}

In Figure~\ref{fig:dens-tde} we compare the inferred circumnuclear density profile surrounding \at\ with those around previous radio-emitting TDEs, as well as to the environments of Sgr A* and M87*.  In all cases we scale the radial profile by the relevant SMBH's Schwarzschild radius.  For \at\ we use the value of $\log{M_{\rm SMBH}}=6.7\pm 0.4$ M$_\odot$, or $R_s=(1.6\pm 0.9)\times 10^{12}$ cm \citep{Cannizzaro2020}. With this value, our observations span a scale of $\approx (5-44)\times 10^3$ $R_s$. 
The density profile is similar to that inferred in ASASSN-14li \citep{Alexander2016}. The inner portion of the sampled profile, $\rho\propto R^{-1.7}$, is roughly consistent with expectations for spherical Bondi accretion, while the final three epochs show evidence for a steepening.

\section{Conclusions}
\label{sec:conclusions}

We presented detailed VLA and ALMA observations of the TDE \at, spanning 55 to 560 days after disruption.  Using these data we inferred the physical properties of the outflow and the circumnuclear environment.  We find that the outflow is non-relativistic ($v\approx 0.07c$) with a total kinetic energy of $\approx 4\times 10^{48}$ erg, typical of previous non-relativistic TDEs. The energy exhibits an initial rise as $E_K\propto t^{1.5}$ to about 200 days, which is likely indicative of continuous energy injection.  The circumnuclear medium has a density of about $1\times 10^4$ cm$^{-3}$ at a radius of $10^{16}$ cm ($6\times 10^3$ $R_s$) and follows a steep radial decline of about $R^{-1.7}$ with a potential further steepening at about $2.5\times 10^4$ $R_s$.

We further find that a mildly relativistic outflow would require an unexpectedly narrow opening angle of $\theta_j\approx 2^\circ$, while a truly relativistic outflow is unphysical.  The ordinary nature of the outflow in \at\ relative to other TDEs and to Type Ib/c SNe (which are more common) casts doubt on the claimed association of this event with a high-energy neutrino.

\software{CASA (McMullin et al. 2007), pwkit (Williams et al. 2017)}

\acknowledgments
We thank Joe Lazio with his assistance with the NE2001 model.  The Berger Time-Domain Group at Harvard is supported by NSF and NASA grants.  K.D.A. is supported by NASA through the NASA Hubble Fellowship grant \#HST-HF2-51403.001-A awarded by the Space Telescope Science Institute, which is operated by the Association of Universities for Research in Astronomy, Inc., for NASA, under contract NAS5-26555. This paper makes use of the following ALMA data: ADS/JAO.ALMA\#2018.1.01766.T. ALMA is a partnership of ESO (representing its member states), NSF (USA) and NINS (Japan), together with NRC (Canada), MOST and ASIAA (Taiwan), and KASI (Republic of Korea), in cooperation with the Republic of Chile. The Joint ALMA Observatory is operated by ESO, AUI/NRAO and NAOJ. The National Radio Astronomy Observatory is a facility of the National Science Foundation operated under cooperative agreement by Associated Universities, Inc.

\appendix

\section{Interstellar Scintillation}
\label{sec:scintillation}

In the observations at days 130 and 161 we find excess emission at about $1.5-3$ GHz.  Here we consider if these are consistent with interstellar scintillation \citep{Goodman1997}, which is known to affect low frequency radio emission from compact sources.  Using the Galactic coordinates of \at\ in the NE2001 electron density model \citep{Cordes2002}, we find a scattering measure of ${\rm SM}\approx 0.34\times 10^{-3}$ kpc-m$^{-20/3}$, and a transition frequency $\nu_{\rm ss}\approx 14$ GHz.  Combining these parameters, we find a scattering screen distance of $d_{\rm scr}\approx 5.64 \nu_{\rm ss}^{-1} (SM_{-3.5})^{6/5}\approx 2.5$ kpc.  The critical radius for diffractive scintillation, which can result in order unity flux variations, is $\theta_{\rm crit}\approx 2.35(SM_{-3.5})^{-3/17}d_{\rm scr, kpc}^{-11/17}\approx 1.3$ $\mu$as.  For \at\ we infer an angular radius of 3.2 $\mu$as at 60 days (and $20$ $\mu$as at 560 days), indicating that diffractive scintillation is not expected. On the other hand, refractive scintillation is relevant on this size scale.  Following \citet{Goodman1997} we expect the peak frequency for refractive scintillation to be $\sim 4$ GHz with a modulation index of $\sim 0.3$, which is consistent with the observed flux variations.

\bibliography{bibliography}
\end{document}